\documentclass[reprint,aps,pra,superscriptaddress]{revtex4-1}
\usepackage{amssymb,amsmath}
\usepackage{graphicx}
\usepackage{dcolumn}
\usepackage{multirow}
\usepackage{color}
\usepackage[english]{babel}
\usepackage[caption=false]{subfig} 

\newcommand{\ket}[1]{\ensuremath{|{#1}\rangle}}

\begin{document}

\def\simlt{\mathrel{\lower .3ex \rlap{$\sim$}\raise .5ex \hbox{$<$}}}

\title{\textbf{\fontfamily{phv}\selectfont 
Quantum control and process tomography of a semiconductor quantum dot hybrid qubit}}
\author{Dohun Kim}
\author{Zhan Shi}
\author{C. B. Simmons}
\author{D. R. Ward}
\author{J. R. Prance}
\author{Teck Seng Koh}
\affiliation{Department of Physics, University of Wisconsin-Madison, Madison, WI 53706}
\author{John King Gamble}
\affiliation{Sandia National Laboratories, Albuquerque, NM 87185, USA}
\author{D. E. Savage}
\author{M. G. Lagally}
\affiliation{Department of Materials Science and Engineering, University of Wisconsin-Madison, Madison, WI 53706, USA}
\author{Mark Friesen}
\author{S. N. Coppersmith}
\author{M. A. Eriksson}
\affiliation{Department of Physics, University of Wisconsin-Madison, Madison, WI 53706}

\maketitle

\textbf{The similarities between gated quantum dots and the transistors in modern microelectronics~\cite{Hanson:2007p1217,Zwanenburg:2013p961} -- in fabrication methods, physical structure, and voltage scales for manipulation -- have led to great interest in the development of quantum bits (qubits) in semiconductor quantum dots~\cite{Loss:1998p120,Hayashi:2003p226804,Petta:2004p1586,Elzerman:2004p431,Petta:2005p2180,Koppens:2006p766,Kouwenhoven:2006p722,Taylor:2007p464,PioroLadriere:2008p776,Foletti:2009p903,Laird:2010p1985,Gaudreau:2011p54,Maune:2012p344,Medford:2013p050501,Medford:2013p654,Dial:2013p146804}.  While quantum dot spin qubits have demonstrated long coherence times, their manipulation is often slower than desired for important future applications, such as factoring~\cite{VanMeter:2008p183}. Further, scalability and manufacturability are enhanced when qubits are as simple as possible.  Previous work has increased the speed of spin qubit rotations by making use of integrated micromagnets~\cite{PioroLadriere:2008p776}, dynamic pumping of nuclear spins~\cite{Foletti:2009p903}, or the addition of a third quantum dot~\cite{Medford:2013p654}.  Here we demonstrate a new qubit that offers both simplicity -- it requires no special preparation and lives in a double quantum dot with no added complexity -- and is very fast: we demonstrate full control on the Bloch sphere with $\pi$-rotation times less than 100~ps in two orthogonal directions. We report full process tomography, extracting high fidelities equal to or greater than 85\% for X-rotations and 94\% for Z-rotations. We discuss a path forward to fidelities better than the threshold for quantum error correction. }

\begin{figure*}
\includegraphics[width=0.8\textwidth]{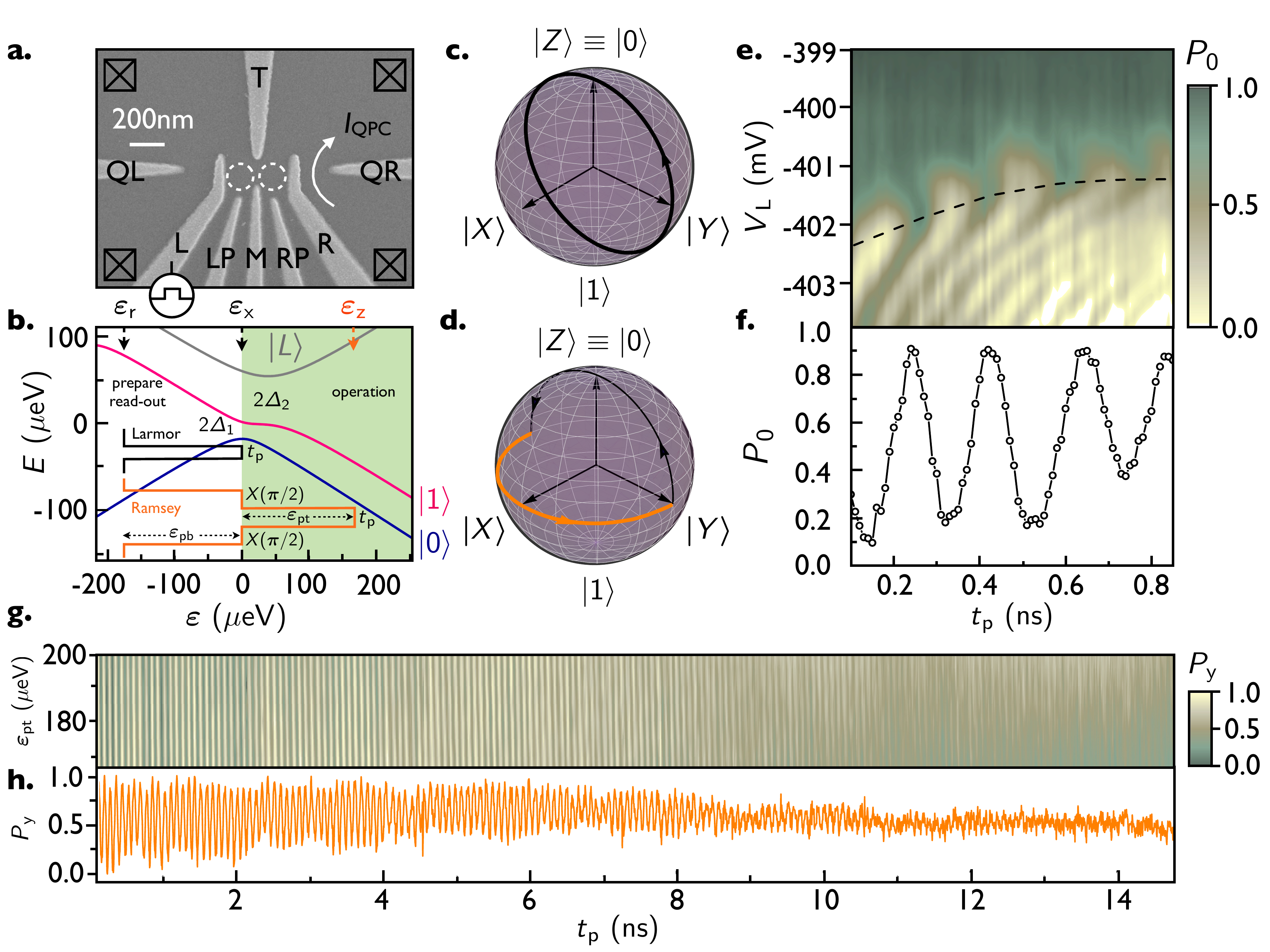}
\caption{
\textbf{Si/SiGe hybrid qubit device, energy levels, and measurement of quantum oscillations. a}, SEM image of a device lithographically identical to the one used in the experiment, with the locations of the double dot shown by white dashed circles. The current through the quantum point contact (QPC) $I_\text{QPC}$ is used for charge sensing via a measurement of its change in the presence of manipulation voltage pulses applied to gate L. 
\textbf{b}, Diagram of the calculated energy levels $E$ versus detuning $\varepsilon$, including the ground states of the $(2,1)$ and $(1,2)$ charge configuration, the singlet-triplet splitting $\delta E$, and the first excited state of the $(1,2)$ charge configuration, with Hamiltonian parameters determined as described in the Supplementary Information. The resulting blue, magenta, and gray solid curves show logical $\ket{0}$, $\ket{1}$, and the primary leakage state $|L\rangle$. Black and orange solid line insets: pulse sequences used for implementation and measurement of rotations about the X-axis (black) and Z-axis (orange). 
\textbf{c-d}, Schematics of the evolution of the Bloch vector during pulse sequences for X-axis (\textbf{c}) and Z-axis (\textbf{d}) rotations.  
\textbf{e}, Probability $P_0$ of the state to be $\ket{0}$, measured for the X-rotation pulse sequence as a function of the voltage $V_\text{L}$ and pulse duration $t_\text{p}$. Dashed curve shows a path along $\varepsilon\simeq 0$~$\mu$eV. 
\textbf{f}, Line-cut of $P_0$ along the dashed curve in \textbf{e}, as a function of $t_\text{p}$, starting at $V_\text{L}\simeq -402$~mV with pulse amplitude $\varepsilon_\text{pb}\simeq 160$~$\mu$eV, showing $\sim 5.2$~GHz quantum oscillations with coherence time $T_2^*\simeq 2$~ns. 
\textbf{g}, Probability $P_\text{y}$ of the state to be in $\ket{Y}=\sqrt{1/2}(\ket{0}+i\ket{1})$, measured for the Z-rotation sequence as a function of pulse amplitude $\varepsilon_\text{pt}$ and pulse duration $t_\text{p}$ of the top pulse. 
\textbf{h}, Line-cut of $P_\text{y}$ as a function of $t_\text{p}$ at $\varepsilon_\text{pt}\simeq 180$~$\mu$eV, showing $\sim 12.1$~GHz oscillations with coherence time $T_2^*\simeq 10$~ns. }
\label{fig1} 
\end{figure*}

Most studies of electron spins in quantum dots have focused on the construction and control of qubits using single~\cite{Elzerman:2004p431,Koppens:2006p766,Kouwenhoven:2006p722,PioroLadriere:2008p776}, double~\cite{Hayashi:2003p226804,Petta:2004p1586,Petta:2005p2180,Taylor:2007p464,Foletti:2009p903,Maune:2012p344,Dial:2013p146804}, and triple~\cite{Laird:2010p1985,Gaudreau:2011p54,Medford:2013p050501,Medford:2013p654} quantum dots, in which each quantum dot is occupied by a single electron.  Here we demonstrate that the quantum dot hybrid qubit  -- a qubit based on three electrons in two dots~\cite{Shi:2012p140503,Koh:2012p250503} -- is different both qualitatively and quantitatively from other quantum dot qubits.  First, the presence of the third electron in the pair of dots enables fast rotations about two independent axes on the qubit Bloch sphere.  Second, because the qubit lives in only two dots rather than three, there is only one relevant control parameter: the energy difference $\varepsilon$ (also called the detuning) between the two quantum dots.  The resulting phase space is only one-dimensional, making it straightforward to find so-called ``sweet spots" where the decoherence effects of charge fluctuations on the qubit are greatly reduced.  We show here that qubit state preparation, full control over two axes on the Bloch sphere, and qubit read-out all can be implemented using fast changes to only a single gate voltage used as the qubit control parameter.

The qubit is formed in a double quantum dot in a Si/SiGe heterostructure~\cite{Thalakulam:2010p183104} (Fig.~1a) with the gate voltages tuned so that two electrons occupy the left dot and one electron occupies the right dot (the $(2,1)$ charge state).  Changing the voltage on gate L to make $\varepsilon$ more positive favors a transition to the $(1,2)$ charge state.  The hybrid qubit consists of the two lowest energy eigenstates with total spin $s = 1/2$ and z-component of spin $s_z = -1/2$~\cite{Shi:2012p140503}.  The one-dimensional phase space parametrized by $\varepsilon$ has an obvious symmetry with respect to the $(2,1)$ and $(1,2)$ states.  By changing $\varepsilon$ adiabatically from positive to negative, we can track the qubit state 
$\ket{0}=\ket{\downarrow}|S\rangle$, 
where $S$ denotes a singlet state in the right dot, from right to left, as shown by the dark blue line in Fig.~1b.  Similarly, the state
$\ket{1}=\sqrt{1/3}\ket{\downarrow}|T_0\rangle+\sqrt{2/3}\ket{\uparrow}|T_-\rangle$, 
where $T_0$ and $T_-$ are two of the triplet states in the right dot, is shown by the magenta line in Fig.~1b. In the $\varepsilon\geq 0$ region of phase space, shaded with a green background in Fig.~1b, the lines showing the energies of states $\ket{0}$ and $\ket{1}$ are nearly parallel, leading to protection from charge noise; this protection is particularly good both near $\varepsilon = 0$ and for $\varepsilon \gg 0$. For moderate negative $\varepsilon$, state $\ket{0}$ has charge occupation $(2,1)$, whereas state $\ket{1}$ has charge occupation $(1,2)$, enabling robust readout through measurement of the current $I_\text{QPC}$ through the charge-sensing quantum point contact (Fig.~1a).  Waiting several tens of nanoseconds in the measurement regime, to enable inelastic tunneling between the $(1,2)$ and $(2,1)$ charge states, re-initializes the qubit to state $\ket{0}$~\cite{Petta:2004p1586}.

\begin{figure*}
\includegraphics[width=0.7\textwidth]{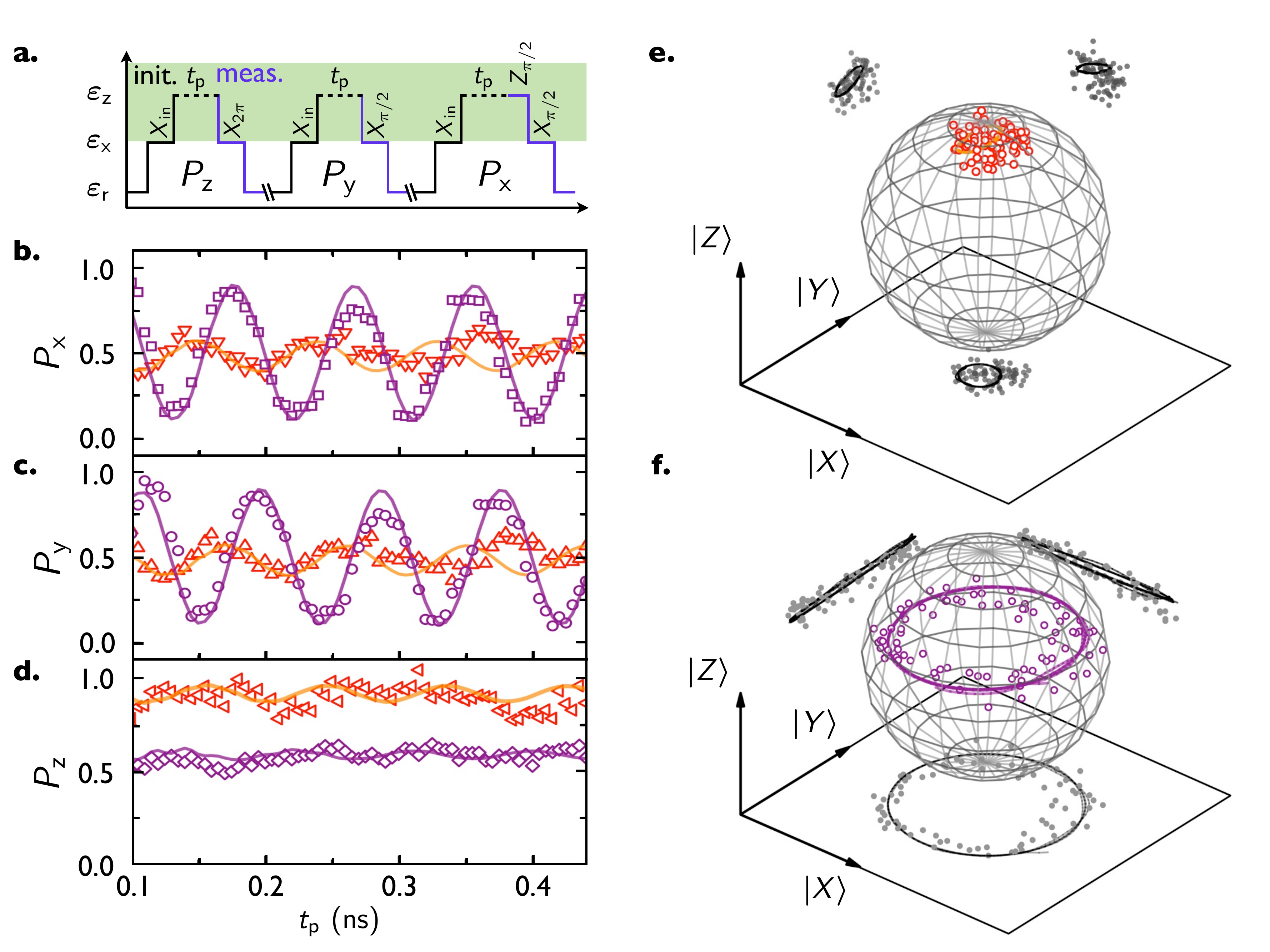}
\caption{
\textbf{State tomography and universal gate control of the hybrid qubit. a}, Schematic diagrams of the pulse sequences used to perform state tomography by measuring the Z-axis projection $P_\text{z}=P_0$, Y-axis projection $P_\text{y}$, and X-axis projection $P_\text{x}$ of the state that is initialized through an $X_\text{in}$-gate and rotated around the Z-axis for an evolution time $t_\text{p}$. 
\textbf{b-d}, $P_\text{x}$, $P_\text{y}$, and $P_\text{z}$ as a function of $t_\text{p}$ for states initialized near $\ket{0}$ (red symbols) and near $\ket{-Y}$ (purple symbols). Solid curves with corresponding colors are calculation from a model using non-adiabatic pulses with 80~ps rise times and the energy level diagram of Fig.~1a, with best fit parameters $2\Delta_1/h= 5.2$~GHz, $2\Delta_2/h= 14.8$~GHz, and $\delta E/h= 12.12$~GHz. The calculations neglect high-frequency dephasing but do include low-frequency ﬂuctuations in the detuning $\varepsilon$~\cite{Petersson:2010p246804,Shi:2013p075416} (See Supplementary Information). 
\textbf{e}, Bloch sphere representation of the measured qubit state evolution under a Z-gate that is initialized near $\ket{0}$. Gray symbols show the XZ, XY, and YZ plane projections of the 3D representation. 
\textbf{f}, Representation with an input state near $\ket{-Y}$. Gray symbols show XZ, XY, and YZ plane projections of the 3D representation. Solid curves show the results of numerical simulations, as described above. The difference between \textbf{e} and \textbf{f} reflects the expected effect of Z-rotations on the two different initial states.}
\label{fig2} 
\end{figure*}

\emph{Demonstration of rotations about two axes on the Bloch sphere.}
As shown schematically in Fig.~1c, X-rotations on the Bloch sphere correspond to oscillations between qubit states $\ket{0}$ and $\ket{1}$.  To demonstrate such oscillations, we first prepare the qubit in state $\ket{0}$, by waiting for initialization at $\varepsilon=\varepsilon_\text{r}$ (see the inset to Fig.~1b).  Changing the detuning abruptly to $\varepsilon_\text{x}$ results in a Hamiltonian $H=\Delta_1\sigma_\text{x}$, where $\sigma_\text{x}$ is the usual Pauli matrix, and the tunnel coupling $\Delta_1$ characterizes the strength of the anticrossing between $\ket{0}$ and $\ket{1}$. Such a Hamiltonian is expected to result in oscillations between states $\ket{0}$ and $\ket{1}$ at the Larmor frequency $2\Delta_1/h$, where $h$ is Planck's constant.  The resulting final state is measured by changing the detuning to $\varepsilon_\text{r}$:  at this detuning, the charge sensing quantum point contact is used to determine whether the charge state is $(2,1)$, corresponding to state $\ket{0}$, or $(1,2)$ corresponding to state $\ket{1}$ (see the Methods Summary and Supplementary Information for details of the measurement procedure). 

Figures~1e and 1f show that this procedure results in rotations about the X-axis of the Bloch sphere. In Fig.~1e we plot the probability $P_0$ of observing state $\ket{0}$ as a function of the pulse duration $t_\text{p}$ and the gate voltage $V_\text{L}$, the latter of which determines $\varepsilon$. The path through Fig.~1e that corresponds to $\varepsilon=0$ is curved (Fig.~1e, dashed curve), because of frequency-dependent attenuation in the microwave coaxial cable~\cite{Petersson:2010p246804}. Fig.~1f shows a line cut through the path corresponding to $\varepsilon=0$, revealing periodic oscillations in $P_0$ at a frequency of 5.2~GHz, corresponding to $\Delta_1/h\simeq 2.6$~GHz.  The visibility of this oscillation is larger than 0.8, and we estimate a lower bound for the coherence time $T_2^*\simeq 2$~ns, by performing an exponential fit to the decay of the first 3~ns of the data.  This X-rotation alone is similar to a charge qubit rotation, and the coherence time is consistent with such an interpretation~\cite{Petersson:2010p246804,Shi:2013p075416,Shi:2014p3020}.  The most important parameter describing a quantum gate is the process fidelity; we return to this quantity below after demonstrating Z-rotations, because two rotation axes are required for process tomography.

Z-rotations of the qubit, shown schematically in Fig.~1d as the orange line about the equator of the Bloch sphere, can be performed by abruptly changing the detuning to a large, positive value,  $\varepsilon_\text{z}>150$~$\mu$eV.  At this detuning, the states $\ket{0}$ and $\ket{1}$ have the same dependence on $\varepsilon$; they differ in energy by a fixed value $\delta E$, the single-triplet splitting of the right dot, and the two states will acquire a relative phase $\varphi=e^{-it_\text{p}\delta E/h}$, where $t_\text{p}$ is the time spent at $\varepsilon_\text{z}$.  Experimentally, the qubit is first prepared in state $\ket{-Y}=\sqrt{1/2}(\ket{0}-i\ket{1})$, by initializing to state $\ket{0}$ and by performing an $\text{X}(\pi/2)$ rotation, as described above.  The resulting state can be rotated about the Z-axis by setting the detuning equal to $\varepsilon_\text{z}$ for a time $t_\text{p}$. We then measure the probability $P_\text{y}$ of being in $\ket{Y}=\sqrt{1/2}(\ket{0}+i\ket{1})$, by applying a second $\text{X}(\pi/2)$ rotation to rotate this state into state $\ket{0}$, and by measuring the charge state at the readout position $\varepsilon_\text{r}$ (see the orange inset to Fig.~1b).  Figures~1g and 1h show the resulting quantum oscillations of the qubit state around the Z-axis of the Bloch sphere.  The oscillations have a visibility larger than 0.85 and reveal a coherence time $T_2^*\simeq 10$~ns, a time that is much longer than the $\text{Z}(\pi/2)$ manipulation time of approximately 25~ps.

\begin{figure*}
\includegraphics[width=0.7\textwidth]{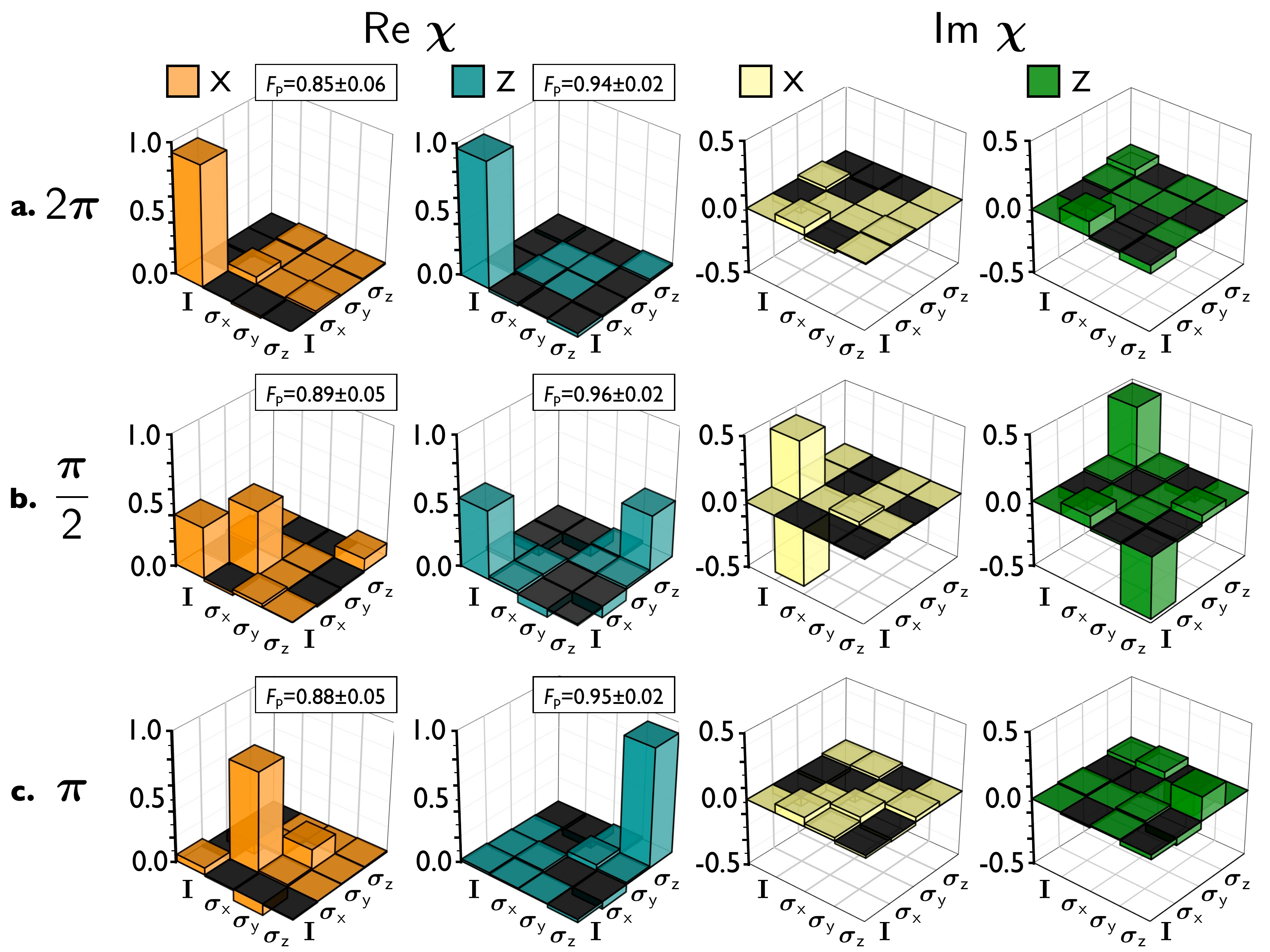}
\caption{
\textbf{Quantum process tomography of the hybrid qubit.} 
Real and imaginary parts of the process matrix $\chi$ obtained by maximum-likelihood estimation for three X and Z-rotation processes in the Pauli basis $\{ \text{I},\sigma_\text{x},\sigma_\text{y},\sigma_\text{z}\}$. 
\textbf{a}, rotation by $2\pi$ (identity),
\textbf{b}, rotation by $\pi/2$, and 
\textbf{c}, rotation by $\pi$. The measured process matrices are close to the targets for each operation; for example, the target for a $\text{X}(\pi/2)$ rotation is $1/2$ for the $(\text{I},\text{I})$ and $(\sigma_\text{x},\sigma_\text{x})$ components, $i/2$ and $-i/2$ for the $(\text{I},\sigma_\text{x})$ and $(\sigma_\text{x},\text{I})$ components, respectively, and zero for all others.  For each process, the average value and uncertainty of the process fidelity $F_\text{p}=\text{Tr}[\chi_\text{ideal}\chi]$ is obtained from $\chi$ for 30 distinct sets of input and output states (see Supplementary Information).}
\label{fig3} 
\end{figure*}

\emph{Universal control of the hybrid qubit and quantum state tomography.}
The initialization, measurement, and two rotation gates just described enable full control of the qubit and tomographic characterization of the resulting qubit state.  Fig.~2a shows schematic diagrams of three pulse sequences that achieve this goal.  For each of the three sequences, the qubit state is prepared in $\ket{0}$ at $\varepsilon_\text{r}$, after which the detuning is pulsed to $\varepsilon_\text{x}$ to perform an X-rotation. By varying the time spent at $\varepsilon_\text{x}$ from 160 to 340~ps -- times that correspond to X-rotation angles from approximately $\pi$ to $3\pi$ -- the qubit at the end of the initalization sequence (``init" in Fig.~2a) can be set to a controlled and nearly arbitrary superposition of $\ket{0}$ and $\ket{1}$.  The limitation on this superposition is set by the X-gate fidelity, which we extract below. In each of the three diagrams, this superposition state then evolves under a Z-gate for a time $t_\text{p}$, reaching nearly the remainder of the Bloch sphere, with limitations again set by the Z-gate fidelity.

Tomographic measurement is then performed by rotating the states $\ket{X}=\sqrt{1/2}(\ket{0}+\ket{1})$, $\ket{Y}=\sqrt{1/2}(\ket{0}+i\ket{1})$, and $\ket{Z}=\ket{0}$ into state $\ket{0}$, through the use of either a single X-gate or a combination of an X and a  Z-gate, as shown by the blue lines in Fig.~2a. The resulting probability $P_0$ is measured by pulsing the detuning back to $\varepsilon_\text{r}$. This initialization and measurement scheme provides universal control of the hybrid qubit and allows us to reconstruct fully the time evolution of the state vector.

As an example of control of the hybrid qubit, we perform Z-rotations on the Bloch sphere starting with two different initial states. Figures 2b-d show the probabilities $P_\text{x}$, $P_\text{y}$, and $P_\text{z}$ of measuring the system in the states $\ket{X}$, $\ket{Y}$, and $\ket{Z}$ following a Z-rotation with the initial state either close to $\ket{0}$ (red data points) or close to $\ket{-Y}$ (purple data points). In Figs.~2e and 2f we plot the Bloch vectors that are extracted from these three measurements for each time $t_\text{p}$. The results obtained are consistent with intuitive expectations: for input state $\ket{0}$, the time evolution under a Z-rotation accumulates a trivial phase, and the Bloch vector remains near the north pole of the Bloch sphere. For input state $\ket{-Y}$, $P_\text{x}$ and $P_\text{y}$ oscillate sinusoidally between nearly $0$ and $1$, whereas $P_\text{z}$ remains roughly constant with magnitude near $1/2$, as expected.

Because the energy level structure of the qubit has been characterized experimentally, and the most likely leakage states are known (see Fig.~1b), the rotation gates can be modeled numerically. The simulations incorporate realistic pulses with rise times of approximately 80~ps. The results of the calculations, shown as the solid orange ($\ket{0}$ initialization) and solid purple ($\ket{-Y}$ initialization) curves in Fig.~2b-f, are in good agreement with the experiment.  As we describe in the Supplementary Information, leakage into state $|L\rangle$ during these pulse sequences is 5\% or less; in future work such leakage could be reduced further by appropriate pulse shaping.  In the Supplementary Information, we also report the results of an analogous state tomography of the qubit evolved under X-rotations, using pulse sequences similar to those shown in Fig.~2a.  

\emph{Quantum process tomography (QPT) of the hybrid qubit.}
For a single qubit the process matrix representation of any output state ${\cal E}(\rho)$ for a given input state $\rho$ resulting from a given quantum process can be written as~\cite{NielsenBook,Chow:2009p090502} 
\begin{equation}
{\cal E}(\rho)=\sum_{m,n=1}^{4}\tilde{E}_m\rho\tilde{E}^\dagger_n\chi_{mn} ,
\end{equation}
where the operators $\tilde{E}_m$ form a basis in the space of $2\times 2$ matrices and $\chi$ is the process matrix. To characterize both rotation axes, here we perform two sets of quantum process tomography: we consider both Z and X-rotations, and for each we characterize rotations by an angle $2\pi$ (nominally equivalent to zero rotation or the identity gate), $\pi/2$, and $\pi$. To determine $\chi_{mn}$ for each process, we prepare four linearly independent input and output states using the manipulation approach presented above and in Fig.~2a. The process matrix is then obtained using maximum-likelihood estimation~\cite{NielsenBook,Chow:2009p090502}. 

Figure~3 shows the results of this procedure; it reports in the Pauli basis $\{\text{I},\sigma_\text{x},\sigma_\text{y},\sigma_\text{z}\}$ the real and imaginary parts of $\chi$ for both X and Z-rotations of magnitude $2\pi$, $\pi/2$, and $\pi$. For each process, we find reasonably good agreement between the estimated process matrix $\chi$ and the ideal process matrix $\chi_\text{ideal}$. The process fidelity $F_\text{p}$ is $F_\text{p}=\text{Tr}[\chi_\text{ideal}\chi]$, yielding $F_\text{p}=0.85\pm 0.06$, $0.89\pm 0.05$, and $0.88\pm 0.05$ for the $\text{X}(2\pi)$, $\text{X}(\pi/2)$, and $\text{X}(\pi)$ processes, respectively, and $0.94\pm 0.02$, $0.96\pm 0.02$, $0.95\pm 0.02$ for the $\text{Z}(2\pi)$, $\text{Z}(\pi/2)$, and $\text{Z}(\pi)$ processes.  The statistical uncertainty is estimated using 30 different sets of input and output states for each process (see Supplementary Information for all the fidelity results for each set of input and output states).

The qubit studied here was formed in natural Si, in which the low density of nuclear spins has sufficient spin protection that spin dephasing does not limit coherence.  Further, the absence of piezoelectric coupling between electrons and phonons eliminates a dephasing channel that would be important in polar semiconductors~\cite{Koh:2013p19695}.  With gate fidelities between 85\% and 95\%, the highest reported to date in an electrically-gated semiconductor quantum dot qubit, the quantum dot hybrid qubit offers a promising combination of characteristics in the past found separately in qubits based on either charge or spin degrees of freedom: a good ratio of manipulation time to coherence time, together with fast overall operation and the ability to fully control the qubit using a single control parameter.  It is important to emphasize the need for fast qubits: although the scaling associated with quantum computation is extremely promising for algorithms such as Shor’s factoring algorithm and Grover’s search algorithm, for the advantages to be realizable on time scales relevant to humans, qubit operations must still be fast~\cite{VanMeter:2008p183}.  The fidelity of both the X and Z-gates reported here appears to be limited currently by the time spent in the regime near $\varepsilon=0$, which is used for X-rotations.  In the future it may be possible to use ac driving of the control parameter to perform X-rotations~\cite{Koh:2013p19695}, so that the regime in which the hybrid qubit is sensitive to charge noise may be avoided entirely.  Even making use of the $\varepsilon=0$ regime for pulsed gating, as we have done here, the hybrid qubit as demonstrated offers a desirable combination of high speed, high fidelity, and efficient control in a semiconductor quantum dot qubit.

\vspace{.2in}
\noindent \textbf{Methods Summary}

\emph{Measurement:} 
The experiments are performed on a double quantum dot fabricated in a Si/SiGe heterostructure~\cite{Thalakulam:2010p183104,Shi:2013p075416,Simmons:2011p156804}. The valence electron occupation is $(2,1)$ or $(1,2)$, where the first (second) number is the charge occupation in the left (right) dot, as confirmed by magnetospectroscopy measurements~\cite{Simmons:2011p156804}. Fast voltage pulses with repetition rate of 15~MHz are generated by Agilent~81134A pulse generator and are added to the dot-defining dc voltage through a bias tee (Picosecond Pulselabs~5546-107) before being  applied to gate L. The conductance change through the quantum point contact (QPC) with and without the manipulation pulses, measured with a lock-in amplifier (EG\&G model~7265), is used to determine the average charge occupation and is converted to the reported probabilities. Charge relaxation during the measurement phase is taken into account using the measured charge relaxation time $T_1\simeq 23.5$~ns at the read-out detuning of $\varepsilon\simeq -160$~$\mu$eV. The coaxial cables leading to the sample at base temperature (electron temperature $\simeq 140$~mK~\cite{Simmons:2011p156804}) in the dilution refrigerator have frequency dependent attenuation that can lead to pulse distortion at the sample. To partially compensate this effect, we developed a pulse correction scheme based on qubit rotation and measurement, which is applied to all the sequences shown in Fig.~2. In the Supplementary Information, we present the details of measurement technique, the probability normalization, and the pulse correction scheme.  

\emph{Theory:} 
Numerical simulations of the experiment were performed based on the energy level diagram in Fig.~1b, using a pulse rise time of 80~ps. Parameters used in the simulation were extracted from experiment, as discussed in the Supplementary Information.  We model the dynamical evolution of the density matrix of the system as a function of detuning and pulse duration $t_\text{p}$ using a master equation $\dot{\rho}=-i[H,\rho ]/\hbar$ with the Hamiltonian $H$ written in a basis of the position eigenstates~\cite{Shi:2013p075416,Shi:2014p3020}. The $(1,2)$ charge occupation probability is extracted at the end of the pulse and is averaged over 2 ns in the measurement stage of the pulse. Low-frequency fluctuations in the detuning $\varepsilon$ are incorporated by convolving the simulation result~\cite{Petersson:2010p246804,Shi:2013p075416} with a Gaussian distribution in $\varepsilon$ of width $\sigma_\varepsilon =5$~$\mu$eV. 

\begin{figure}[b]
\includegraphics[width=3in]{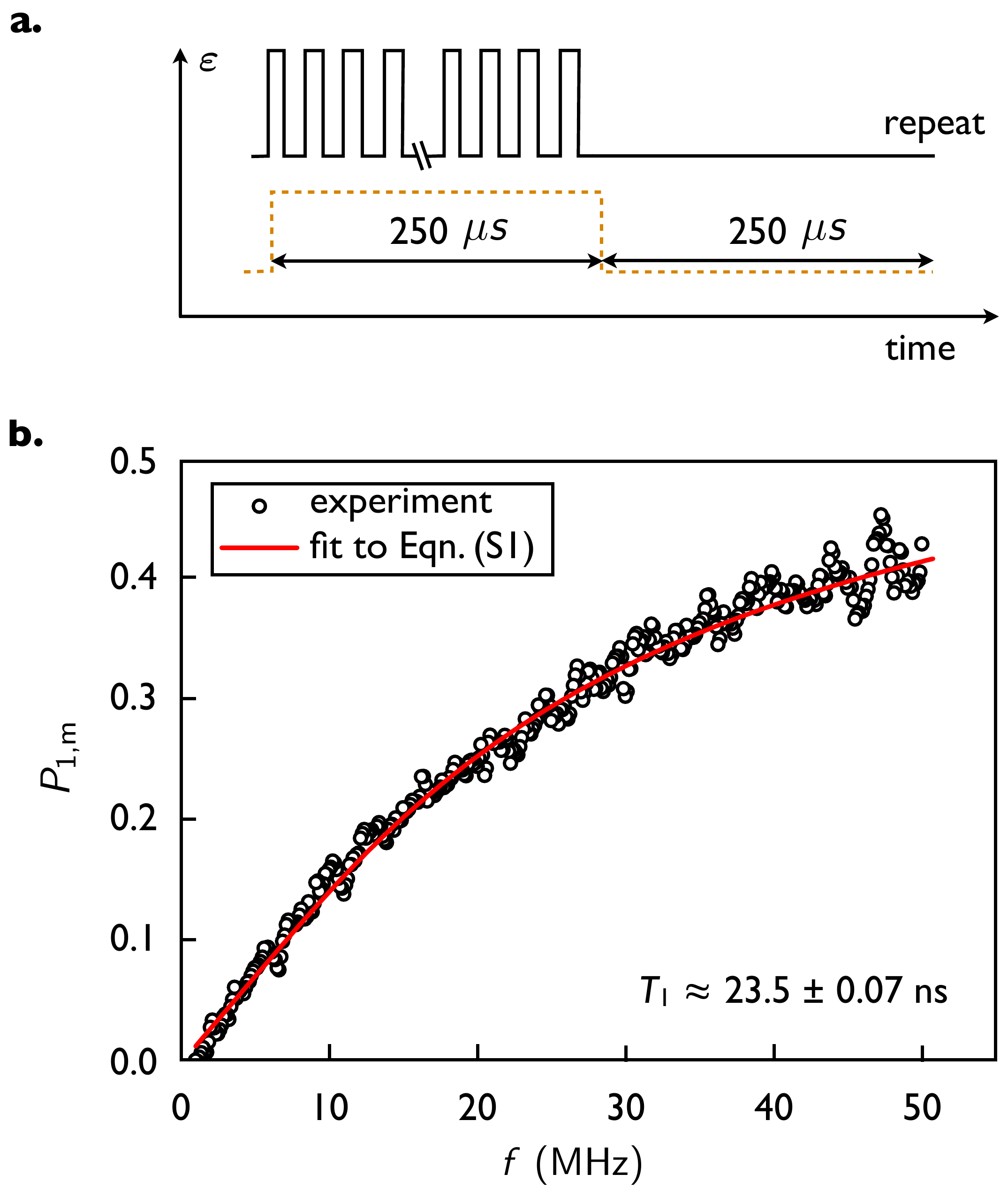}
\caption{
\textbf{Measurement details. a}, Pulse sequence used for lock-in measurement of the qubit state. The orange dashed line shows the corresponding lock-in reference signal, which also serves as the probability calibration pulse. 
\textbf{b}, Measurement of charge relaxation time $T_1$:  probability $P_{1,m}$ of being in state $\ket{1}$ at read-out detuning $\varepsilon_\text{r}=-160$~$\mu$eV after an $\text{X}(5\pi)$ rotation as a function of manipulation pulse repetition rate $f$.  The probability $P_{1,m}$ decreases as $f$ is decreased because of charge relaxation with charge relaxation time $T_1$.  The red solid curve shows a fit to Eq.~(\ref{eq:S1}) with best fit parameter $T_1\simeq 23.5$~ns.  }
\label{fig:S1} 
\end{figure}

\emph{Acknowledgements}
This work was supported in part by ARO (W911NF-12-0607) and by the Laboratory Directed Research and Development program at Sandia National Laboratories.  Sandia National Laboratories is a multi-program laboratory managed and operated by Sandia Corporation, a wholly owned subsidiary of Lockheed Martin Corporation, for the U.S. Department of Energy's National Nuclear Security Administration under contract DE-AC04-94AL85000.  Development and maintenance of the growth facilities used for fabricating samples is supported by DOE (DE-FG02-03ER46028). This research utilized NSF-supported shared facilities at the University of Wisconsin-Madison. DK acknowledges useful conversations with Xian Wu and Kenneth Rudinger.

\emph{Author Contributions}
MAE and SNC conceived the experiment. DK developed pulse sequences for qubit operation and tomography, performed electrical measurements and numerical simulations with the aid of ZS, and analyzed the data with MAE and SNC. CBS fabricated the quantum dot device. JRP and DRW developed hardware and software for the measurements. TSK, JKG, and MF aided with the theoretical analysis. DES and MGL prepared the Si/SiGe heterostructure. DK, SNC, and MAE wrote the manuscript with the contributions of all authors. 
 
\emph{Additional Information}
The authors declare no competing financial interests. Supplementary information accompanies this paper. Correspondence and requests for materials should be addressed to Mark A. Eriksson (maeriksson\emph{@}wisc.edu).

\appendix

\begin{figure}[t]
\includegraphics[width=3.in]{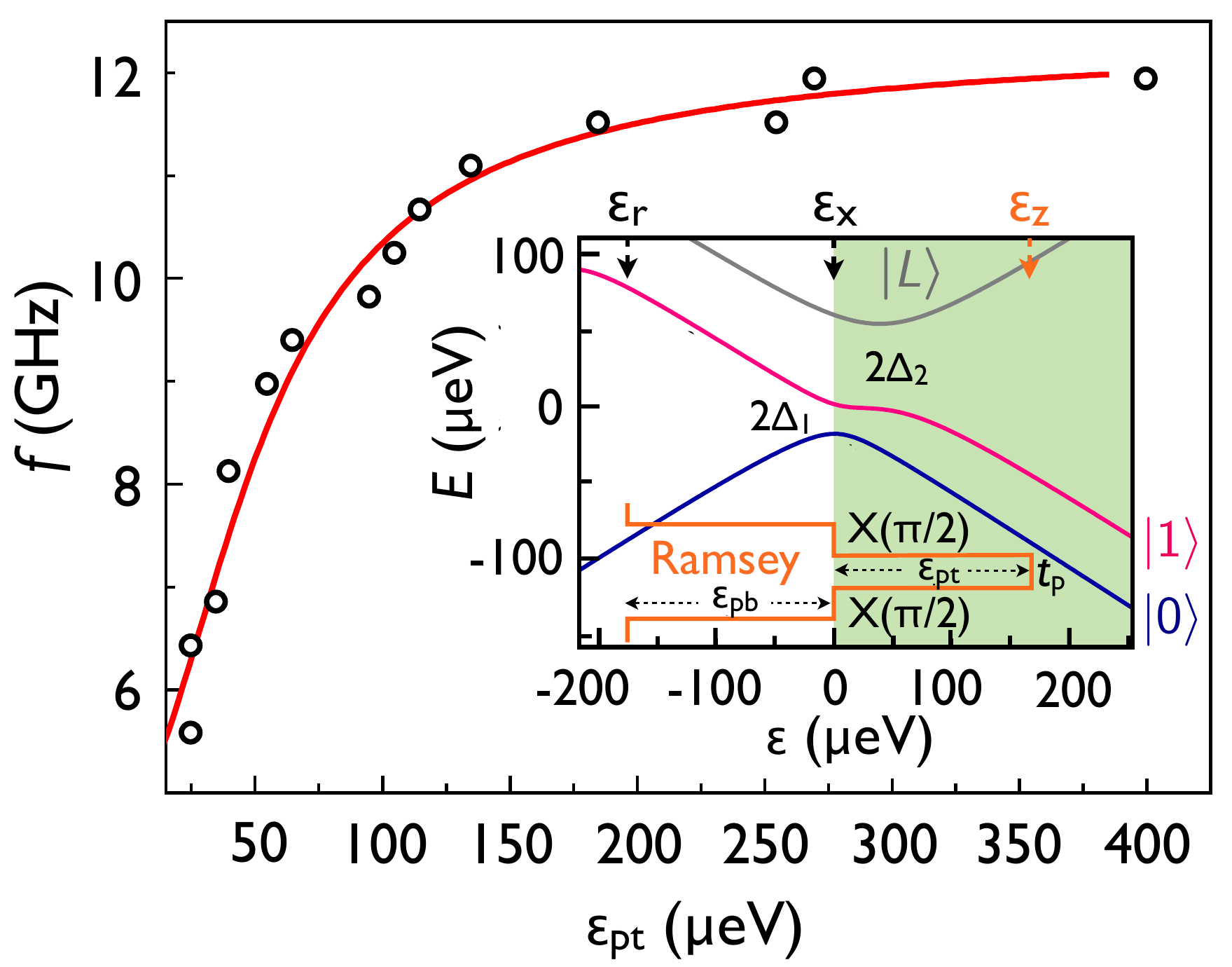}
\caption{
\textbf{Energy level spectroscopy to determine the qubit Hamiltonian parameters.}
Oscillation frequency $f$ of Ramsey fringes as a function of top pulse amplitude $\varepsilon_\text{pt}$ with fixed bottom pulse amplitude $\varepsilon_\text{pb}\simeq 160$~$\mu$eV. The red solid curve shows a fit to the model described in Appendix~\ref{sec:S3}. The inset shows the energy level diagram with the pulse sequence used for the measurement.  The frequency $f$ corresponds to the energy difference between the two lowest energy states.}
\label{fig:S2} 
\end{figure}

\section{QPC operation and measurement details}
The quantum point contact (QPC) is operated at a fixed voltage bias of 0.5~mV, and its current is measured with a current preamplifier (DL Instruments model 1211). For the measurement of changes in charge occupation probabilities resulting from fast manipulation pulses, we use the general approach shown in Fig.~\ref{fig:S1}a. We alternate an appropriate number of manipulation pulses at a repetition rate of 15~MHz with a time period of no pulses of the same duration, in order to form a low frequency signal with frequency $\sim 2$~kHz. Pulses are generated using an Agilent Technologies 81134A pulse generator with a 10-90\% rise time of approximately 80~ps. The data are acquired using a lock-in amplifier with a reference signal corresponding to the presence and absence of the pulses, as shown schematically by the orange dashed line in Fig.~\ref{fig:S1}a.  This technique allows us to measure the difference between the QPC conductance with and without the manipulation pulse train. We compare the measured signal level with the corresponding $(2,1)$-$(1,2)$ charge transition signal level, calibrated by sweeping gate L and applying the orange square pulse shown in Fig.~\ref{fig:S1}a to gate L with an amplitude the same as the manipulation pulses. The resulting lock-in signal represents the rms change of the QPC conductance corresponding to a one-electron transition. The background lock-in signal corresponding to state $\ket{0}$ is determined by using the same manipulation pulse sequences but moving the readout detuning $\varepsilon_\text{r}$ to deep within the $(2,1)$ charge ground state, where the manipulation pulses do not induce $(2,1)$-$(1,2)$ charge transitions. 

\begin{figure*}
\includegraphics[width=0.75\textwidth]{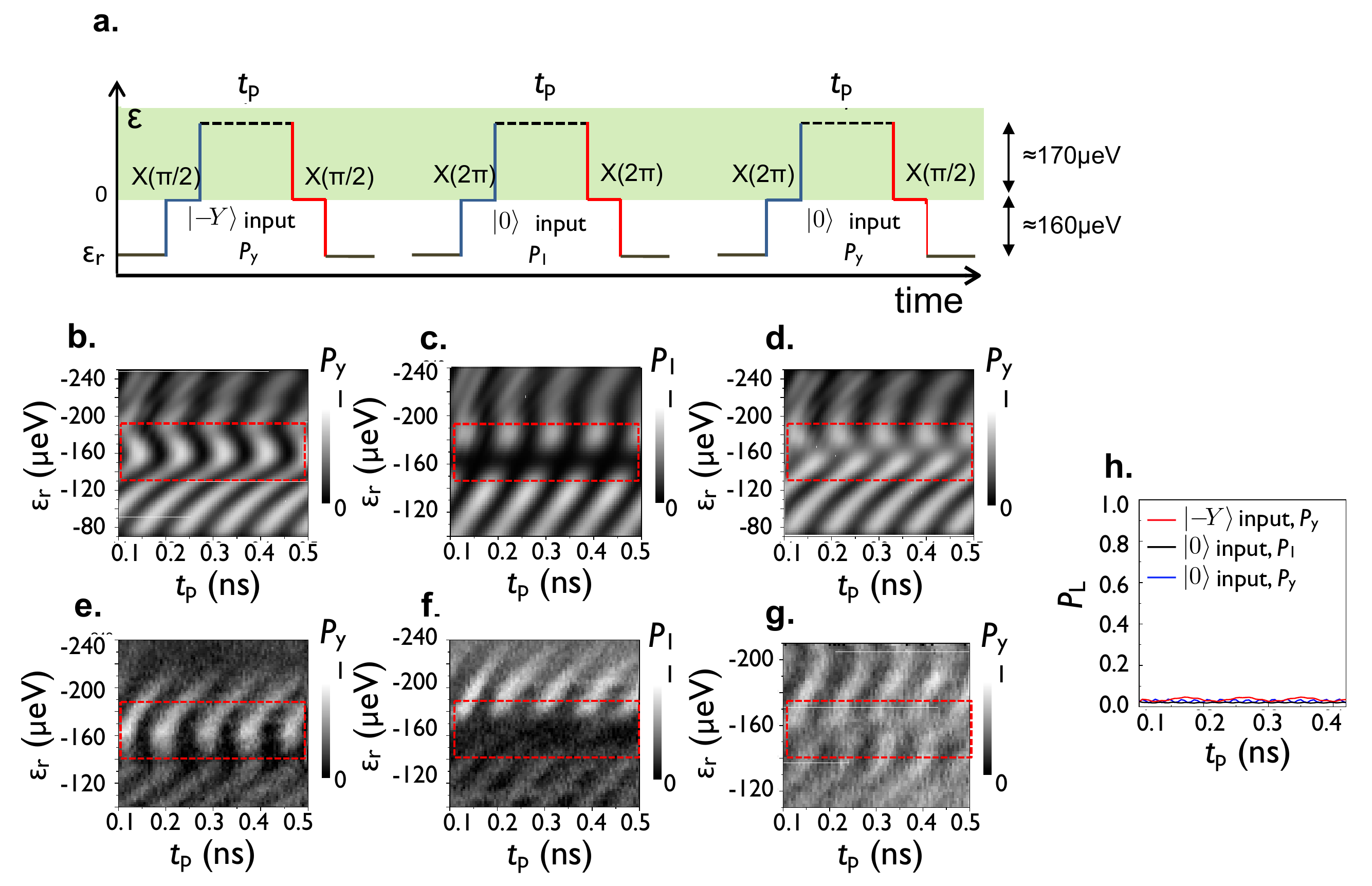}
\caption{
\textbf{Numerical simulation, comparison with experiment, and leakage estimation. a}, Schematic pulse sequences for numerical simulation and experiment shown in 
\textbf{b-g} using various input states and measurement axes determined by control X-gates with amplitude of 160~$\mu$eV relative to $\varepsilon_\text{r}$ and time evolution under Z-gate for time $t_\text{p}$. 
\textbf{b-d}, Numerical simulation results for \textbf{b}, $\ket{-Y}$ input and Y-axis projection $P_\text{y}$, 
\textbf{c}, $\ket{0}$ input and Z-axis projection $P_1$, and 
\textbf{d}, $\ket{0}$ input and Y-axis projection $P_\text{y}$ as a function of read out detuning $\varepsilon_\text{r}$ and $t_\text{p}$.  Dashed rectangles show regions near $\varepsilon_\text{r}\simeq -160$~$\mu$eV where the voltage step corresponding to the X-gate reaches $\varepsilon=0$, the detuning where the X-gate duration corresponds to the intended rotations desired for initialization and measurement. 
\textbf{e-g}, Corresponding experimental results for \textbf{e}, $\ket{-Y}$ input and $P_\text{y}$ measurement, 
\textbf{f}, $\ket{0}$ input and $P_1$ measurement, and 
\textbf{g}, $\ket{0}$ input and $P_\text{y}$ measurement as a function of $\varepsilon_\text{r}$ and $t_\text{p}$. 
\textbf{h}, Calculated leakage state probability $P_\text{L}$ at $\varepsilon_\text{r}=-160$~$\mu$eV after applying pulse sequences shown in \textbf{a}.}
\label{fig:S3} 
\end{figure*}

Finally, we calibrate to account for charge relaxation during the measurement phase. To estimate the charge relaxation time $T_1$ at $\varepsilon_\text{r}$ we measure the uncalibrated probability $P_{1,m}$ of state $\ket{1}$ after an $\text{X}(5\pi)$ evolution, which will depend on the pulse repetition rate. Figure~\ref{fig:S1}b shows $P_{1,m}$ as a function of pulse repetition rate, showing the expected reduction of signal size arising from charge relaxation. We fit the data to the form~\cite{Petta:2004p1586,Wang:2013p046801}
\begin{equation}
\label{eq:S1}
P_{1,m}=P_1\frac{T_1(1-e^{-\tau/T_1})}{\tau} ,
\end{equation}
where $\tau=1/f$ is the measurement time and $P_1$ is the state $\ket{1}$ probability without charge relaxation (that is, the probability immediately after the rotation pulse ends).  From Eq.~(\ref{eq:S1}) we obtain $T_1\simeq 23.5\pm 0.07$~ns. Using this $T_1$ time allows us to correct for charge relaxation by appropriate normalization of the data to account for relaxation from state $\ket{1}$ to state $\ket{0}$ during the measurement process.

\section{Energy level spectroscopy} \label{sec:S2}
We characterize the energy levels of the qubit using a Ramsey fringe measurement pulse sequence roughly corresponding to X, Z, and X-rotations (see the inset of Fig.~\ref{fig:S2}). We fix the base pulse amplitude $\varepsilon_\text{pb}\simeq 160$~$\mu$eV and measure Ramsey fringes as a function of the peak pulse amplitude $\varepsilon_\text{pt}$. Figure~\ref{fig:S2} shows the oscillation frequency determined from a Fourier transform of the experimental data as a function of $\varepsilon_\text{pt}$. We fit the data to the four-state model described below in Appendix~\ref{sec:S3} to obtain the energy difference between states $\ket{0}$ and $\ket{1}$ as a function of detuning (see the inset of Fig.~\ref{fig:S3}). We fix the first anti-crossing energy $\Delta_1/h\simeq 2.6$~GHz, determined from the X-rotation (Larmor oscillation) measurement shown in Fig.~1 of the main text, and we extract the second anti-crossing energy  $\Delta_2/h\simeq 7.5 \pm 0.2$~GHz and singlet-triplet energy splitting $\delta E/h\simeq 12.1 \pm 0.2$~GHz from the fit, shown as the red line in Fig.~\ref{fig:S2}.  

\begin{figure}[t]
\includegraphics[width=3.2in]{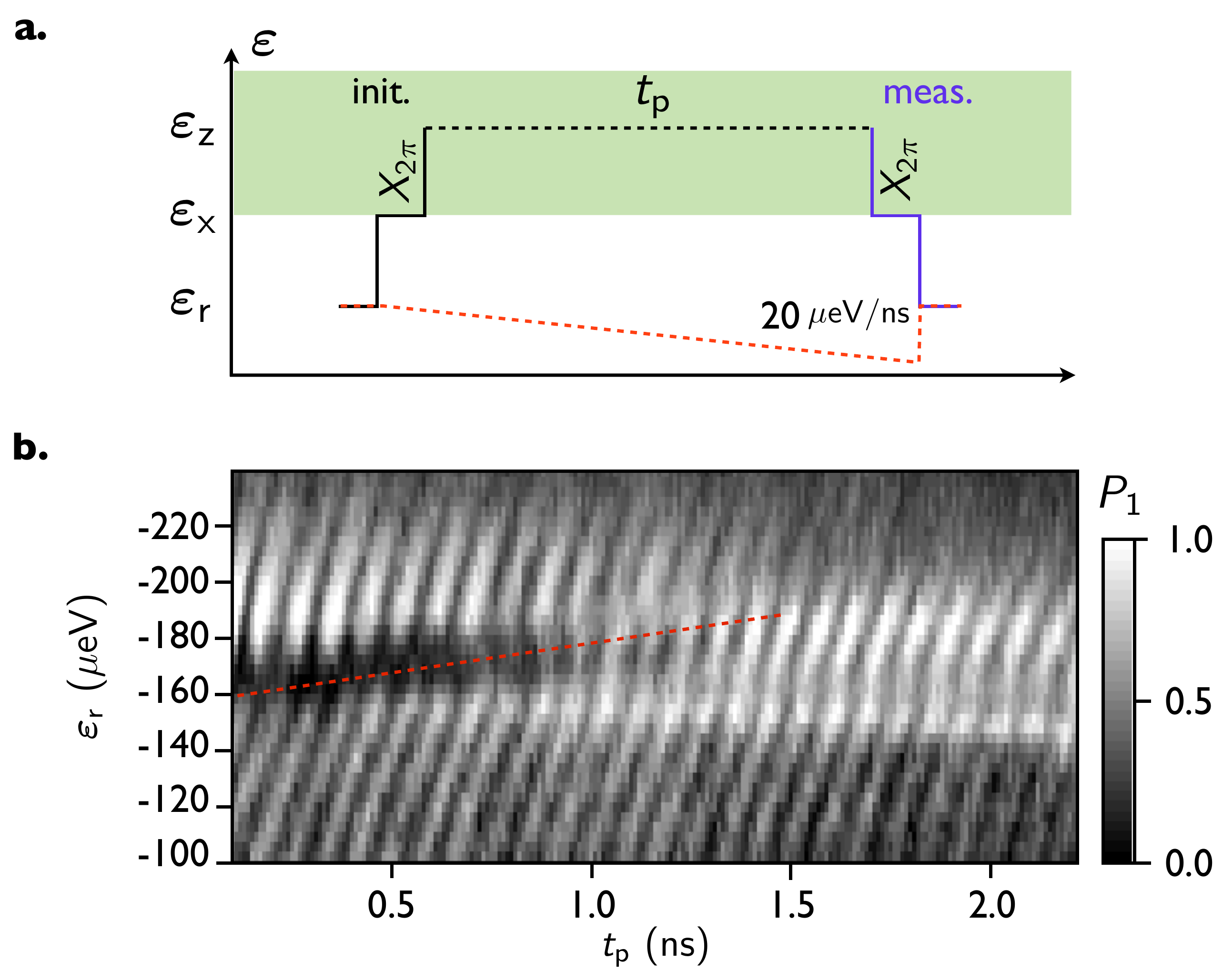}
\caption{
\textbf{Pulse correction scheme. a}, Schematic pulse sequence used for estimation of the pulse correction needed to correct deviations caused by frequency-dependent attenuation in the coaxial lines. The red dashed pulse shows a linear ramping correction estimated from \textbf{b}. 
\textbf{b}, Probability $P_1$ of the time evolved state to be in state $\ket{1}$ as a function of evolution time $t_\text{p}$, measured with no ramping correction. At $\varepsilon_\text{r}\simeq -160$~$\mu$eV, the input and measurement axis control step is performed by $\text{X}(2\pi )$ at $\varepsilon = 0$, which ideally produces $P_1 = 0$. The red dashed line shows an increase of the level mismatch between the pulse voltages corresponding to input and measurement axis control X-gates.  This mismatch generally produces $P_1 > 0$ near $\varepsilon_\text{r}\simeq -160$~$\mu$eV. From the slope of the mismatch growth, a linear ramping correction of 20~$\mu$eV/ns is estimated and applied to all the tomographic measurements in this work.}
\label{fig:S4} 
\end{figure}

\section{Numerical simulation and leakage state population} \label{sec:S3}
Here we discuss the comparison of experimental results with numerical simulations that demonstrate that the quantum oscillation data are consistent with the measured Hamiltonian parameters, which can be used to characterize leakage out of the qubit subspace. We model the dynamical evolution of the density matrix $\rho$ of the three electron, double quantum dot system under the realistic pulses having rise time approximately 80~ps by a master equation~\cite{Shi:2013p075416,Shi:2014p3020}
\begin{equation}
\label{eq:S2}
\dot{\rho}=-\frac{i}{\hbar}[H,\rho] ,
\end{equation}
where $H$ is the four-state Hamiltonian that we write in the basis of the ground and first excited states of the electron charge states $(2,1)$ and $(1,2)$: $\{ (2,1)_\text{g} , (2,1)_\text{e}, (1,2)_\text{g}, (1,2)_\text{e}\}$.  In this basis, $H$ is given by 
\begin{equation}
\label{eq:S3}
H=
\begin{pmatrix}
\varepsilon/2 & 0 & \Delta_1 & -\Delta_2 \\
0 & \varepsilon/2+\delta E_\text{L} & -\Delta_3 & \Delta_4 \\
\Delta_1 & -\Delta_3 & -\varepsilon/2 & 0 \\
-\Delta_2 & \Delta_4 & 0 & -\varepsilon/2+\delta E
\end{pmatrix} .
\end{equation}
Here, $\Delta_{1-4}$ are tunnel coupling matrix elements, $\varepsilon$ is the detuning, and $\delta E_\text{L}$ and $\delta E$ are the energy separation of the ground and excited $(2,1)$ and $(1,2)$ states, respectively. The parameters $\Delta_3$, $\Delta_4$, and $\delta E_\text{L}$ are relevant to high energy leakage states which we do not experimentally access in this work, and we use the values determined from our previous study~\cite{Shi:2014p3020}. The values of the tunnel couplings $\Delta_1$ and $\Delta_2$, and the singlet-triplet energy separation $\delta E$ are determined as described in Appendix~\ref{sec:S2}. For the simulations reported in this paper, the full four-state Hamiltonian is used.  In some other situations it might be useful to use an effective two-state Hamiltonian, as derived in Ref.~\cite{Ferraro:2014p1}.  We note that state $(2,1)_\text{e}$ plays essentially no role in this paper, because it is outside the detuning and energy ranges relevant to this work; it is included for completeness.

\begin{figure*}
\includegraphics[width=0.7\textwidth]{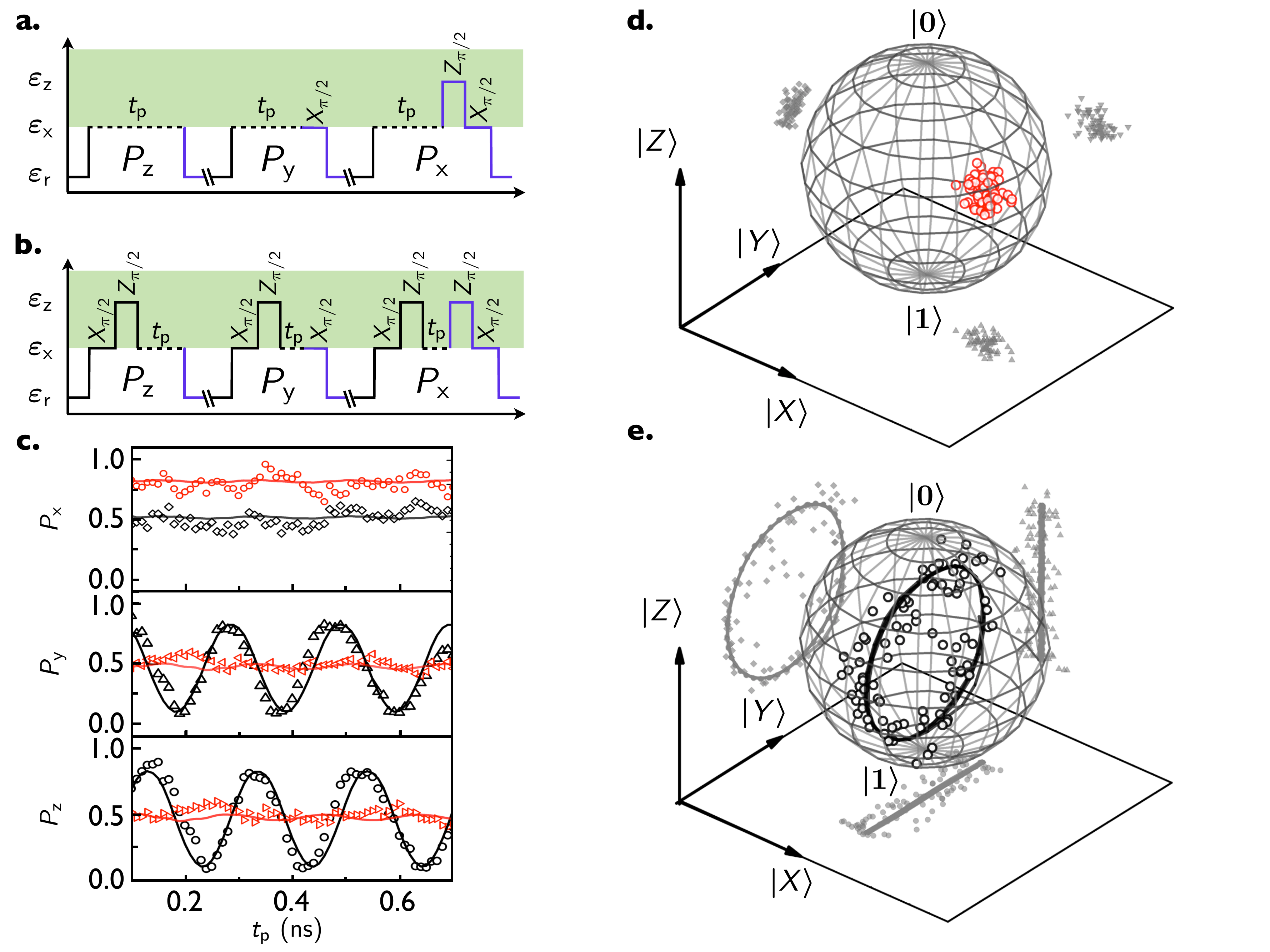}
\caption{
\textbf{State tomography of the states evolved under an X-gate.}
Pulse sequences for state initialization, time evolution, and measurement axis control for state tomography with 
\textbf{a}, input state $\ket{0}$ and 
\textbf{b}, input state $\ket{X}=\sqrt{1/2}(\ket{0}+\ket{1})$. 
\textbf{c}, X-axis projection $P_\text{x}$, Y-axis projection $P_\text{y}$, Z-axis projection $P_\text{z}= P_0$ of the state that is rotated around the X-axis for an evolution time $t_\text{p}$. Black symbols: initial state near $\ket{0}$, Red symbols: initial state near $\ket{X}$. The solid curves with corresponding colors are fits to the model described in Appendix~\ref{sec:S3}. 
\textbf{d}, Bloch sphere representation of the qubit state evolution under an X-gate after initialization near $\ket{X}$. Gray symbols show XZ, XY, and YZ plane projections of the 3D representation. 
\textbf{e}, Bloch sphere representation with initialization state $\ket{0}$. Gray symbols show XZ , XY, and YZ plane projections of the 3D representation. Solid curves show the results of the numerical simulation described above. }
\label{fig:S5} 
\end{figure*}

In the simulations the $(1,2)$ charge occupation probability is extracted at the end of the pulse and is averaged over 2~ns in the measurement stage of the pulse. Low-frequency fluctuations in the detuning $\varepsilon$ are incorporated by convolving the simulation result with a Gaussian distribution in $\varepsilon$ of width $\sigma_\varepsilon = 5$~$\mu$eV~\cite{Petersson:2010p246804,Shi:2013p075416}. We performed a comparison between simulation and experiment for all of our tomographic pulse sequences. As a representative example, Fig.~\ref{fig:S3} shows a comparison of numerical simulations and experiments for Z-gate tomographic measurements as a function of the read-out detuning $\varepsilon_\text{r}$ and the Z-gate evolution time $t_\text{p}$. As shown in Figs.~\ref{fig:S3}b-d (simulation) and e-g (experiment), one can clearly observe that the intended input state, the desired time evolution under a Z-rotation gate, and control of the measurement axis control all can be achieved when $\varepsilon_\text{r}\simeq -160$~$\mu$eV, where the amplitude of the voltage step corresponding to the X-gate reaches $\varepsilon\simeq 0$~$\mu$eV (red dashed rectangles in Figs.~\ref{fig:S3}b-g). For example, for input state $\ket{-Y}$, $P_\text{y}$ oscillates sinusoidally between 0 and 1 (see Figs.~\ref{fig:S3}b and e) while $P_1$ is roughly constant at $1/2$ (see Figs.~\ref{fig:S3}d and g).  Outside the optimal control region, the background probability oscillates both in experiment and in the simulations, because of Landau-Zener-Stuckelberg interference effects~\cite{Shi:2013p075416,Petta:2010p669,Ribeiro:2013p086804}, which arise because the edges of the voltage pulse pass through the anticrossing with a rise time of approximately 80~ps, which is neither fully abrupt nor fully adiabatic.

As with any qubit, leakage is possible out of the three-electron qubit presented here. From the inset to Fig.~\ref{fig:S2}, we can see that state $\ket{L}$ is the most likely leakage state: the charge sensing measurements of the $(1,2)$ charge occupation cannot distinguish $\ket{1}$ from $\ket{L}$ at $\varepsilon_\text{r}$.  We estimate the leakage state population by calculating the excited state charge occupation $P_\text{L}=P(1,2)_\text{e}$ at the end of the pulse sequence. Figure~\ref{fig:S3}f shows the estimated leakage state population $P_\text{L}$ as a function of pulse duration. We find that $P_\text{L}$ is $\sim 3$ to 5\% and does not increase significantly as a function of $t_\text{p}$. 

\section{Pulse optimization} \label{sec:S4}
In the high-frequency pulsed gating that we report here, deviations from an ideal result occur most likely due to imperfections in the low temperature, high frequency setup, such as frequency-dependent attenuation and reflection~\cite{Foletti:2009p903,Petersson:2010p246804}. Although it is difficult to determine the exact transfer function of the transmission line from the room temperature electronics to the qubit, we developed a measurement scheme that enables a first order correction to the applied voltage pulses. 

\begin{figure*}
\includegraphics[width=0.6\textwidth]{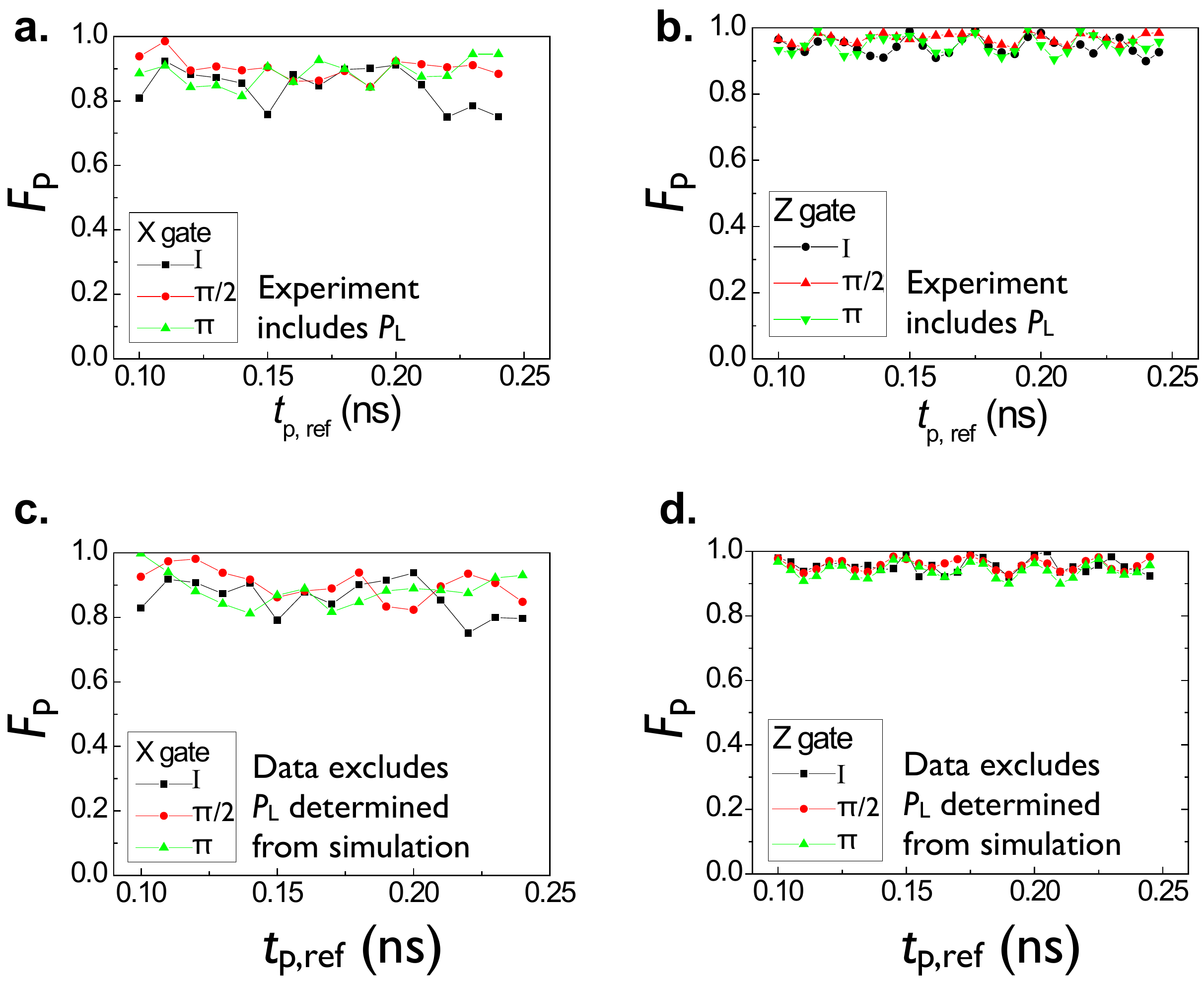}
\caption{
\textbf{Statistics of process fidelities.}
Process fidelity $F_\text{p}= \text{Tr}[\chi_\text{ideal}\chi ]$ of the identity (black square), $\pi/2$ (red circle), and $\pi$ (green triangle) processes performed by the 
\textbf{a}, X and 
\textbf{b}, Z-gates using the experimentally measured input and output states at reference time $t_{\text{p,ref}}$. Results for the averages and standard deviations of the process fidelities are shown in Fig.~3 of the main text.  For the results shown in \textbf{a} and \textbf{b}, no attempt is made to partition the measured $(1,2)$ probability into components corresponding to the qubit state $\ket{0}$ or the leakage state $\ket{L}$.  To check the significance of the leakage, panels \textbf{c} and \textbf{d} report $F_\text{p}$ for \textbf{c}, X and \textbf{d}, Z-gates where the fraction of the probability $P_\text{L}$ corresponding to state $\ket{L}$ is excluded. $P_\text{L}$ is estimated using the simulation described in Appendix~\ref{sec:S3}. As is clear from the graphs, leakage during the gate operation does not significantly affect the extracted gate fidelity.}
\label{fig:S6} 
\end{figure*}

Figure~\ref{fig:S4}a shows a pulse sequence used for a particular (and typical) measurement of state tomography: it is intended to initialize into state $\ket{0}$ and perform a $P_1$ measurement after a Z-axis rotation. As shown in Appendix~\ref{sec:S3} (see Figs.~\ref{fig:S3}c and f), this pulse sequence ideally produces zero probability of being in state $\ket{1}$ near $\varepsilon_\text{r}\simeq -160$~$\mu$eV. One expects that any deviation $\delta \theta$ of the time-evolved Bloch vector from the ideal ($\theta = 0$) will be reflected in the measurement and be proportional to $P_1=1-\cos^2 \delta\theta$. Figure~\ref{fig:S4}b shows the measurement result as a function of $\varepsilon_\text{r}$ and $t_\text{p}$, showing near-ideal results for short $t_\text{p}$, whereas we find a strong deviation from the ideal expectation as the total pulse length exceeds 0.5~ns. We ascribe this effect to an increasing level mismatch caused by pulse imperfections between the voltages corresponding to the initialization rotation and measurement axis control, both of which should be $\text{X}(2\pi )$ gates. From the measured rate of increase of the deviation as a function of pulse duration -- which can be see to be about 20~$\mu$eV/ns, from the superimposed red dashed line in Fig.~\ref{fig:S4}b -- we approximate the correction to be a linear ramp with opposite slope, shown as the red dashed line in Fig.~\ref{fig:S4}a. For pulse durations longer than those used here, the needed correction saturates.  Because we work with pulse durations shorter than this time, we applied the simple linear ramp to all the pulse sequences used for tomographic measurements in this work.  

\section{State tomography of X-rotations}
In addition to the state tomography measurements presented in the main text, we also performed tomography of the qubit states time-evolved under X-gates. Figures~\ref{fig:S5}a and b show schematic diagrams of pulse sequences used for controlling the input states and measurement axes. For desired input state $\ket{0}$, we prepare the initial state $\ket{0}$ at $\varepsilon_\text{r}$ and pulse to $\varepsilon_\text{x} = 0$, where the state rotates about the X-axis of the Bloch sphere. We also prepare input state $\ket{X}$ by applying first an $\text{X}(\pi/2 )$ rotation to state $\ket{0}$, followed by a $\text{Z}(\pi/2 )$ rotation. Similar to the measurement axis control scheme in the main text, we use $\text{X}(0\pi )$, $\text{X}(\pi/2)$, and $\text{Z}(\pi/2)$ followed by $\text{X}(\pi/2)$ gate operations, respectively, in order to project the Z, Y, and X-axes to the Z-axis.  The resulting probability is measured by pulsing back to $\varepsilon_\text{r}$. 

Figures~\ref{fig:S5}c-e show the probability $P_\text{x}$ of X-axis projection $P_\text{x}=P(\sqrt{1/2}(\ket{0}+\ket{1}))$, $P_\text{y}$ of Y-axis projection $P_\text{y}=P(\sqrt{1/2}(\ket{0}+i\ket{1}))$, and $P_\text{z}$ of Z-axis projection $P_\text{z}=P_0$ of the states evolved at $\varepsilon_\text{x} = 0$~$\mu$eV under an X-gate and their Bloch sphere representations with (1) initialization close to $\ket{0}$ (black) and (2) close to $\ket{X}$ (red). The results are consistent with intuitive expectations: for an input state $\ket{X}$, the time evolution under an X-gate acquires a trivial phase only. For input state $\ket{0}$, $P_\text{z}$ and $P_\text{y}$ oscillate sinusoidally between approximately 0 and 1, while $P_\text{x}$ is roughly constant at $1/2$. In Figs.~\ref{fig:S5}c-e, the solid curves show numerical simulations using the model described in Appendix~\ref{sec:S3}, which is in good agreement with the experiment. Similar to the state tomography under a Z-gate, fast dephasing (that is dephasing during a given pulse sequence) is neglected in the simulations, but slow charge noise (that is, detuning noise slower than the pulse repetition rate) is incorporated by convolving results with a Gaussian distribution in $\varepsilon$ of width $\sigma_\varepsilon= 5$~$\mu$eV [23,24]. This set of state tomography data allows us to estimate X-gate process fidelities and to perform quantum process tomography, as we discuss in detail in Appendix~\ref{sec:S6}.  

\section{Quantum process tomography (QPT)} \label{sec:S6}
Performing state tomography for different input states enables us to perform quantum process tomography (QPT) for both the X and Z-gates.  For a given set of operations, the process $\cal{E}(\rho)$ can be written as~\cite{NielsenBook,Chow:2009p090502}
\begin{equation} \label{eq:S4}
{\cal E}(\rho)=\sum_{m,n=1}^{4}\tilde{E}_m\rho\tilde{E}^\dagger_n\chi_{mn} ,
\end{equation}
where $\chi_{mn}$ is known as the process matrix and is defined with respect to an operator basis, which we take as the Pauli basis $\tilde{E}=\{\text{I},\sigma_\text{x},\sigma_\text{y},\sigma_\text{z}\}$.  By definition, the $\chi$ matrix must be Hermitian and must have eigenvalues bounded between 0 and 1. We restrict $\chi$ to be physical by employing maximum-likelihood estimation~\cite{Chow:2009p090502,Poyatos:1997p390}. We first define a strictly physical $\chi$ by constraining it to have the form
\begin{equation}
\chi=\frac{T^\dagger T}{\text{Tr}[T^\dagger T]} ,
\label{eq:S5}
\end{equation}
where $T$ is a lower-triangular matrix in the form
\begin{equation}
\label{eq:S6}
T=
\begin{pmatrix}
t_1 & 0 & 0 & 0 \\
t_5+it_6 & t_2 & 0 & 0 \\
t_{11}+it_{12} & t_7+it_8 & t_3 & 0 \\
t_{15}+it_{16} & t_{13}+it_{14} & t_9+it_{10} & t_4
\end{pmatrix} .
\end{equation}
The form (\ref{eq:S6}), which has 16 real parameters, ensures that $\chi$ is Hermitian and positive semi-definite~\cite{Chow:2009p090502,Poyatos:1997p390}. To determine $\chi_{mn}$ for a given process, we select four linearly independent input states and their corresponding output states for a given process, as determined from the state tomography data (Figs.~2b-d in the main text for Z-gates, and Fig.~\ref{fig:S5}c for X-gates). Next, the measured data is fit to a physical state by numerically minimizing the function
\begin{widetext}
\begin{equation}
f(\mathbf{t})=\sum_{j,k=1}^2 \left[ \rho_{\text{out},jk}- \left( 
\sum_{m,n=1}^{4}\tilde{E}_m\rho_\text{in}\tilde{E}^\dagger_n(T^\dagger T/\text{Tr}[T^\dagger T])_{mn}
\right)_{jk} \right]^2
\label{eq:S7}
\end{equation}
\end{widetext}
where $\mathbf{t}=\{t_1,t_2,\dots,t_{16}\}$, and $\rho_\text{out}$ and $\rho_\text{in}$ are the measured output and input density matrices for a given process. For the initial guess of the $\chi$ matrix, we use the result of a linear inversion of Eq.~(\ref{eq:S4}). From the resultant $T$ matrix, we obtain the most probable completely positive $\chi$ matrix corresponding to the measured values, according to Eq.~(\ref{eq:S5}). We perform QPT on two sets of three processes that each consist of Z and X-gates: (1) $\{ \text{Z}(2\pi)=\text{I}, \text{Z}(\pi/2), \text{Z}(\pi)\}$ and  (2) $\{ \text{X}(2\pi)=\text{I}, \text{X}(\pi/2), \text{X}(\pi)\}$.

To estimate uncertainties in the process fidelity estimation, $F_\text{p}= \text{Tr}[\chi_\text{ideal}\chi ]$, we repeated the maximum-likelihood estimation for a given process using input and output states that are chosen at 30 different reference times $t_\text{p,ref}$ from the state tomography data. Figure~\ref{fig:S6} shows the $F_\text{p}$ obtained from maximum-likelihood estimation as a function of $t_{\text{p,ref}}$ in steps of 10~ps (5~ps) for X (Z)-gate processes, respectively. From the statistical mean and standard deviation of the resulting estimates, we obtain process fidelities of $0.85\pm 0.06$, $0.89\pm 0.05$, and $0.88\pm 0.05$ for $\text{X}(2\pi)=\text{I}$, $\text{X}(\pi/2)$, and $\text{X}(\pi)$ processes, respectively, and $0.94\pm 0.02$, $0.96\pm 0.02$, $0.95\pm 0.02$ for $\text{Z}(2\pi)=\text{I}$, $\text{Z}(\pi/2)$, and $\text{Z}(\pi)$ processes. We compare the $F_\text{p}$ estimation using the experimental data (which includes $P_\text{L}$), shown in Figs.~\ref{fig:S6}a and b, with the case of excluding $P_\text{L}$, as determined from the simulations, as shown in Figs.~\ref{fig:S6}c and d.  The results show that leakage effects are small enough that the fidelity estimates do not depend on whether or not the leakage probability arising during the qubit rotations is excluded or not.  The 3 to 5\% leakage reported above in Fig.~\ref{fig:S3}h thus appears to arise predominately from the initialization and measurement processes. 

\bibliography{siliconqcsnc}

\end{document}